# Magnetic order and its interplay with structure phase transition in van der Waals ferromagnet VI$_3$

Yiqing Hao[1#], Yiqing Gu[1#], Yimeng Gu[1], Erxi Feng[2], Huibo Cao[2], Songxue Chi[2], Hua Wu[3] and Jun Zhao[1,4,5*]

[1]State Key Laboratory of Surface Physics and Department of Physics, Fudan University, Shanghai 200433, China

[2]Neutron Scattering Division, Oak Ridge National Laboratory, Oak Ridge, Tennessee 37831, USA

[3]Laboratory for Computational Physical Sciences (MOE), State Key Laboratory of Surface Physics and Department of Physics, Fudan University, Shanghai 200433, China

[4] Institute of Nanoelectronics and Quantum Computing, Fudan University, Shanghai 200433, China

[5]Shanghai Qi Zhi Institute, Shanghai 200232, China

Abstract

Van der Waals magnet VI$_3$ demonstrates intriguing magnetic properties that render it great for use in various applications. However, its microscopic magnetic structure has not been determined yet. Here, we report neutron diffraction and susceptibility measurements in VI$_3$ that revealed a ferromagnetic order with the moment direction tilted from the *c*-axis by ~36° at 4 K. A spin reorientation accompanied by a structure distortion within the honeycomb plane is observed at a temperature of ~27 K, before the magnetic order completely disappears at $T_C$ = 50 K. The refined magnetic moment of ~1.3 $\mu_B$ at 4 K is considerably lower than the fully ordered spin moment of 2 $\mu_B$/V$^{3+}$, suggesting the presence of a considerable orbital moment antiparallel to the spin moment and strong spin-orbit coupling in VI$_3$. This results in strong magnetoelastic interactions that make the magnetic properties of VI$_3$ easily tunable via strain and pressure.

Elucidating the nature of magnetism in low dimensional materials is important in fundamental and applied physics [1]. The recent discovery of ferromagnetism in two-dimensional (2D) van der Waals materials down to the monolayer limit [2-4] has triggered tremendous research interest. The 2D nature of the van der Waals magnets enables their magnetism to be tuned by various external stimuli [2-13], rendering them highly promising to be used in spintronic devices [14-20]. In the heavily studied 2D ferromagnetic semiconductor CrI$_3$, the spin degree of freedom largely accounts for the material's magnetism. Therefore, various tuning methods focus on the spin degree of freedom [2,5-12]. Recently, VI$_3$, a semiconducting van der Waals ferromagnet, has received considerable interest in research [21-30]. Although both CrI$_3$ and VI$_3$ belong to the family of transition metal trihalides, the magnetic ions have different electron configurations ($3d^3$ for Cr$^{3+}$, and $3d^2$ for V$^{3+}$). While the orbital moment of Cr$^{3+}$ is largely quenched in the octahedral crystal field, V$^{3+}$ has three degenerate orbitals ($L$ = 1). Therefore the orbital degree of freedom could play a more important role in VI$_3$ than in CrI$_3$ ($L$ = 0) [31-34].

Experimental studies showed that the lattice, magnetic, and electronic structures of VI$_3$ [21-30] are more complicated compared to those of CrI$_3$. In contrast to the trigonal low-temperature phase observed in CrI$_3$ [35], VI$_3$ exhibits two-step structure phase transitions at $T_{s1}$ = 79 K and $T_{s2}$ = 32 K [21-25]. Different experimental studies have reported different low-temperature lattice structures in VI$_3$, including trigonal (*R*-3) [21], monoclinic (*C*2/*c*) [23,24], and triclinic (*P*-1) [25] space groups. The V-V bonds lengths are thus found in hexagon [21], anti-dimerization [23] and dimerization [25] for different structures. While the magnetic moments in CrI$_3$ are aligned with the *c*-axis throughout the ferromagnetic phase [35,36], certain experiments suggest that the magnetization easy axis in VI$_3$ deviates from the *c*-axis [27,28]. The magnetic moment magnitude also varies in different measurements. For instance, the magnetic moments along the *c*-axis are found to range between 0.95 and 2.47 $\mu_B$/V$^{3+}$ [21-23,27-29]. Unlike CrI$_3$ that is fully ordered (~3 $\mu_B$/Cr$^{3+}$) [35,36], the deviation of V$^{3+}$ magnetic moment from classical estimations is possibly due to the large orbital magnetic moment and strong spin-orbit coupling that VI$_3$ displays [31-34]. However, there have been contradicting opinions regarding the character of the orbital moments. Certain authors suggested that the orbital magnetic moment is parallel to the spin magnetic moment [34], while others instead suggested they are antiparallel-aligned [31-33]. It was also proposed that the twin orbital orders would result in zero orbital moments [37].

In this work, we use neutron diffraction and thermodynamic measurements to investigate the structural and magnetic ordering properties of VI$_3$. Our polycrystalline samples were grown using the solid-state reaction method. A stoichiometric mixture of vanadium powder and iodine grains was sealed in evacuated quartz tubes, which were flushed through Ar gas. The sealed quartz tubes were first heated to 120 °C for 4 h; then annealed at 400 °C for 2 days. The resulting powder was grounded and placed in the Ar atmosphere for 2 days. The VI$_3$ single crystals were grown using the chemical vapor transport method [23]. Single-crystal neutron diffraction experiments were performed on the HB-3A Four-Circle Diffractometer equipped with a 2D detector at the High Flux Isotope Reactor (HFIR), Oak Ridge National Laboratory (ORNL). Powder neutron diffraction experiment was carried out on the HB-3 Triple-Axis Spectrometer at HFIR, ORNL. 4.5 grams of VI$_3$ powder sample was loaded and sealed in an aluminum can in He glove box. Pyrolytic graphite filters were used to eliminate higher order neutrons. Temperature-dependent magnetic susceptibility and specific heat data were collected using Quantum Design DYNACOOL system. Field-dependent magnetization measurements were performed using Quantum Design MPMS3 SQUID magnetometer.

VI$_3$ consists of vanadium honeycomb layers coupled via van der Waals interactions. The I$^-$ ions are octahedrally coordinated around each V$^{3+}$ ion (Fig. 1a). Field-dependent magnetization data of VI$_3$ at 2 K show clear ferromagnetic hysteresis in both the *c*-axis and *ab*-plane (Fig. 1b), contrary to neglectable coercivity in CrI$_3$ at the same temperature [35]. Temperature-dependent magnetization and specific heat data clearly show two transitions at $T_C$ = 50 K and $T_{s1}$ = 79 K (Fig. 1c,d), corresponding to the ferromagnetic transition and structural phase transition, respectively.

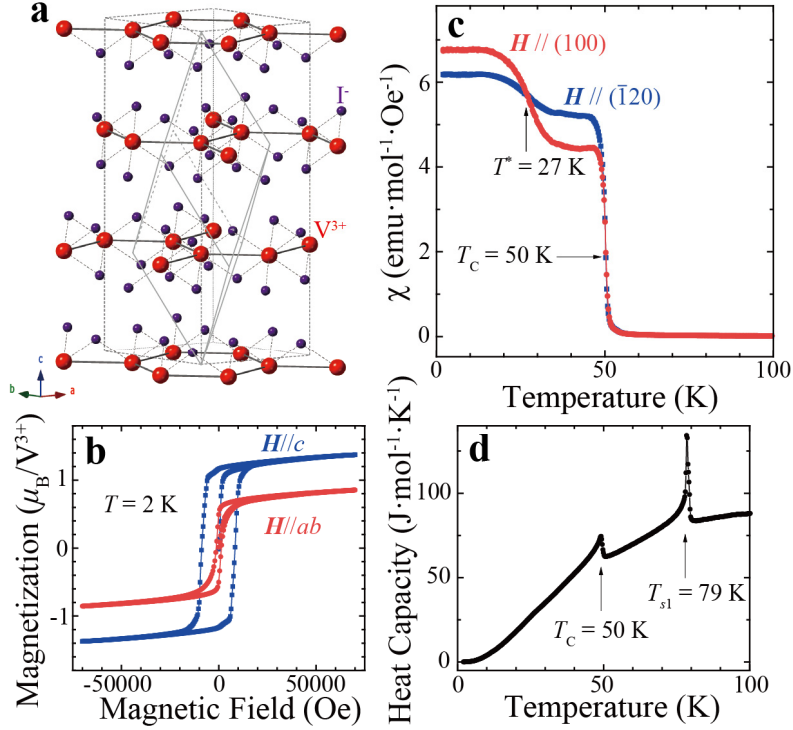

Fig. 1. (a) Crystal structure of VI$_3$ represented in a trigonal lattice. The rhombohedral lattice is also drawn by solid and dashed gray lines in the center. (b) Field-dependent magnetization in *c*-axis (blue) and *ab*-plane (red) at 2 K showing clear ferromagnetic hysteresis loops. (c) Temperature-dependent field-cooled magnetic susceptibilities. Magnetic field of $H$ = 200 Oe was applied along the hexagonal (100) (red) and its orthogonal direction within the *ab*-plane (blue); $T^*$ represents the proposed transition temperature of spin reorientation. (d) Temperature-dependent specific heat of VI$_3$. Two peaks are found at $T_C$ = 50 K and $T_{s1}$ = 79 K.

To determine the lattice and magnetic structure, we performed neutron powder diffraction (NPD) measurements on VI$_3$ polycrystalline samples. The results are summarized in Fig. 2, which illustrates the NPD patterns at 6, 40 and 60 K. Figure 2 insets show the subtracted magnetic intensity at 6 and 40 K to the paramagnetic state at 60 K. Clear magnetic Bragg peaks are observed at 6 and 40 K at $\bm{Q}$ = (003)$_h$, (101)$_h$, (012)$_h$ and (104)$_h$. Here, lower index *h* denotes that the peaks are indexed to the hexagonal notation of the room-temperature *R*-3 trigonal space group. The magnetic Bragg peaks are characterized using the Rietveld refinements. We find the magnetic intensities of both 40 and 6 K are best fitted by a ferromagnetic structure, where the magnetic easy axis is tilted from the *c*-axis toward the *ab*-plane. At 40 K, the *ab*-plane and the *c*-axis moments are 0.50(3) $\mu_B$ and 0.78(6) $\mu_B$, respectively, and at 6 K, the *ab*-plane and *c*-axis moments are 0.69(2) $\mu_B$ and 1.03(4) $\mu_B$, respectively. Intriguingly, the angle between the magnetic easy axis and honeycomb *c*-axis is nearly the same as the temperature drops from 40 K to 6 K. These results differ from CrI$_3$, where the Cr$^{3+}$ magnetic moments are ferromagnetically aligned with the *c*-axis [35,36].

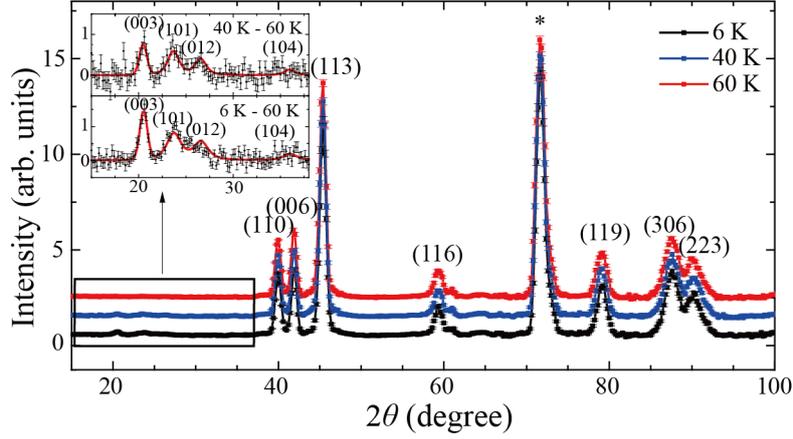

Fig. 2. Neutron power diffraction pattern of VI$_3$. The wavelength used is 2.359 Å. Black, blue, and red lines represent diffraction intensities measured at 6, 40 and 60 K, respectively. Diffraction patterns for 40 and 60 K are shifted in the *y*-axis for clarity. The magnetic peaks and main nuclear Bragg peaks are indexed using hexagonal notation. The star denotes the background signal from the aluminum sample can. Insets show the Rietveld refinements of the subtracted magnetic peaks at 40 and 6 K using the magnetic structure described in the main text. The black dots and red lines represent the observed and calculated intensities, respectively.

We further performed neutron diffraction measurements on a high quality single-crystalline VI$_3$ at the HB-3A Four-Circle Diffractometer. The results are illustrated in Fig. 3. Figure 3a shows the longitudinal cuts of (300)$_h$ peaks at 60 K. Clear differences in the peak positions are observed in (300)$_h$ and (030)$_h$ peaks. These results suggest a C$_3$ symmetry breaking, which has also been noticed in previous low-temperature X-ray diffraction measurements [23-25]. Similar splitting in peak positions are also observed at (306)$_h$, (036)$_h$ and (3-36)$_h$ at 4 K (Fig. 3b). The lattice parameters are refined from fitting more than ten Bragg peaks along different directions at each temperature (Table I). We show that the V-V bonds are dimerized below $T_s$, where the V-V bond length along one direction ((100)$_h$ in the hexagonal notation) is shortened by ~1% compared to the other two V-V bond lengths (Fig. 3g). The splitting of (300)$_h$ peaks as well as the dimerized V-V bond lengths match with recently reported X-ray diffraction measurements [25]. Magnetic peaks at (003)$_h$, (101)$_h$ and (012)$_h$ are measured at 4, 40 and 60 K, respectively, and are illustrated in Fig. 3c-e. Clear enhancement of the above magnetic peaks is observed at 4 and 40 K compared to 60 K, which agrees with our powder measurements. It should be noted the residue intensity of (003)$_h$ at 60 K is due to the $\lambda/2$ contamination of the strong Bragg peak of (006)$_h$. The temperature dependence of the (003)$_h$ peak revealed a clear transition at $T_C$ (Fig. 3f). Critical exponential fitting yields $T_C$ = 50.3(5) K and critical exponent $\beta$ = 0.19(2), which is consistent with our heat capacity measurements (Fig. 1d).

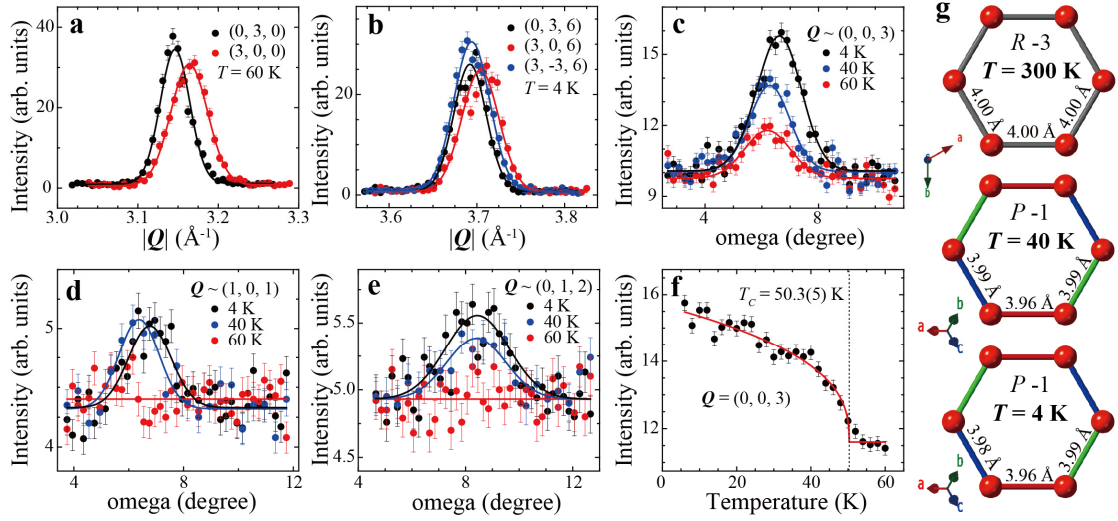

Fig. 3. Nuclear and magnetic Bragg peaks measured in single-crystalline VI$_3$. The wavelength used is 1.542 Å. (a) (300)$_h$ and (030)$_h$ measured at 60 K. (b) (306)$_h$, (036)$_h$ and (3-36)$_h$ measured at 4 K. Black, red and blue colors represent Bragg peaks along different in-plane directions. (c-e) (003)$_h$, (101)$_h$ and (012)$_h$ magnetic peaks measured at indicated temperatures. Solid lines are Gaussian fitting curves based on the magnetic structure refinements. (f) Temperature dependence of (003)$_h$ magnetic Bragg peak. The red line shows fitting using the critical exponential function. (g) V-V bond lengths at different temperatures. Red, blue and green colors indicate three types of symmetric inequivalent V-V bonds below $T_{s1}$. The arrows indicate the lattice directions in hexagonal (upper) and triclinic (middle and lower) notations.

We note that, while the intensity of (003)$_h$ drastically increases when the temperature drops from 40 K to 4 K (Fig. 3c), the intensity of (101)$_h$ remains nearly unchanged (Fig. 3d). Such peak intensity redistributions indicate the presence of a spin reorientation in the honeycomb $ab$-plane between 4 and 40 K. Indeed, our refinements suggest that the magnetic moments at 4 K are $\bm{m}_{//} = 0.84(5) \mu_B$ and $\bm{m}_c = 1.08(8) \mu_B$, with the in-plane moment parallel to the (100)$_h$ direction (Fig. 4a, c, e). At 40 K, the in-plane moment is rotated by 90° and becomes perpendicular to (100)$_h$ (Fig. 4b, d, f). The magnetic moment magnitudes at 40 K are $\bm{m}_\perp = 0.51(5) \mu_B$, $\bm{m}_c = 0.65(6) \mu_B$. To further confirm the presence of spin reorientation within the honeycomb $ab$-plane, we measured the temperature and angle dependence of the magnetization within the $ab$-plane (Fig. 1c). Below $T^* = 27$ K, the magnetization along (100)$_h$ is greater than the magnetization perpendicular to (100)$_h$, and the magnitude relation is reversed above $T^*$, as is expected for the spin reorientation transition revealed from our neutron diffraction experiment. This transition temperature $T^*$ coincides with the $T_{FM} \approx 26$ K revealed from previous magnetization measurements [28,30]. Nevertheless, the angle of the in-plane moment rotation is 90° via our single crystal refinement rather than 30° reported by previous angular dependent magnetization measurements [28].

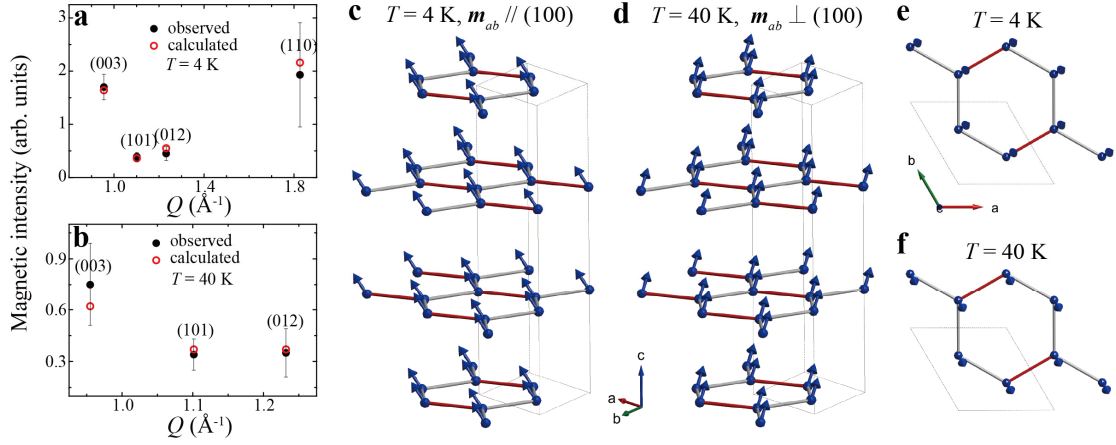

Fig. 4. (a, b) Magnetic structure refinement of $VI_3$ at 4 and 40 K. Black dots and red circles represent observed and calculated magnetic peak intensities, respectively. (c, d) Magnetic structures of $VI_3$ at 4 and 40 K. Blue arrows represent moment directions of $V^{3+}$ and red bonds represent shortened V-V bonds. Gray lines represent a unit cell in hexagonal notation. (e, f) Projections of the 4 and 40 K magnetic structures in honeycomb $ab$-plane.

Using powder and single crystal neutron diffraction refinements, we estimate the ordered moment of $VI_3$ to be ~1.3(1) $\mu_B/V^{3+}$. In contrast to $CrI_3$, where the $Cr^{3+}$ moments are fully ordered [35,36], this value is much smaller than the classic estimation of pure spin moment of 2 $\mu_B/V^{3+}$. The reduced ordered moment in $VI_3$ can be understood through the orbital magnetic moment, which is unneglectable, aligned nearly antiparallel to the spin moment [31-33]. The spin-orbit coupling induces strong magnetic anisotropy, which may easily stabilize the magnetic order down to the monolayer limit. Meanwhile, the in-plane spin orientations of $VI_3$ strongly depend on local [$VI_6$] octahedral environments. Therefore they can be easily tuned via structural distortions. As a result, in-plane spin reorientation is observed at $T^* = 27$ K accompanied by a structural distortion within the honeycomb plane [25]. The magnetism of $VI_3$ can be largely tuned via strain and pressure. This calls for further research, focusing especially on the material's few-layer form.

In summary, we conducted neutron diffraction and susceptibility measurements and showed that the ferromagnetic moment of the van der Waals magnet $VI_3$ is tilted from the $c$-axis by ~36° at 4 K. By increasing the temperature to ~27 K, the in-plane moments rotate by 90° from parallel to perpendicular to the $(100)_h$, which is accompanied by a shrinking V-V bond distance. The magnetic moments are ~0.93(8) $\mu_B$ and 1.24(5) $\mu_B$ at 40 and 6 K, respectively, and are much lower compared to the fully ordered spin magnetic moment of $V^{3+}$. The magnetic moment reduction and the magnetoelastic coupling can be understood by considering the presence of a considerable orbital magnetic moment and strong spin-orbit coupling, which result in strong magnetoelastic interactions and are useful for designing highly tunable spintronic devices.

This work was partially supported by the Innovation Program of Shanghai Municipal Education Commission (Grant No. 2017-01-07-00-07-E00018), the Shanghai Municipal Science and Technology Major Project (Grant No. 2019SHZDZX01), and the National Natural Science Foundation of China (Grant No. 11874119). E.F. and H.C. acknowledge the support of

|   | 60K | 40K | 4K |
|---|---|---|---|
| a | 7.685(11) | 7.676(8) | 7.674(11) |
| b | 7.702(11) | 7.699(11) | 7.699(14) |
| c | 7.711(11) | 7.706(11) | 7.695(19) |
| α | 53.46(10) | 53.44(10) | 53.33(15) |
| β | 53.11(10) | 53.12(10) | 53.07(14) |
| γ | 53.22(9) | 53.20(9) | 53.22(11) |
| V-V(1) | 3.967(7) | 3.964(7) | 3.963(9) |
| V-V(2) | 3.991(10) | 3.989(10) | 3.977(15) |
| V-V(3) | 3.996(8) | 3.992(8) | 3.988(10) |

Table I. Lattice parameters and V–V bond lengths calculated from single-crystal diffraction. The space group is *P*-1 (No. 2), and the lattice axes are based on room-temperature rhombohedral lattice coordinates as shown in Fig. 1a.

*Email: zhaoj@fudan.edu.cn